\documentclass[a4paper,11pt]{article}
\usepackage{jinstpub} 
\usepackage{lineno}
\usepackage{url}

\usepackage{amsmath}
\usepackage{siunitx}

\usepackage{subcaption}
\usepackage{makecell}

\title{A Cost Effective Optimization of the hybrid-DOM Design for TRIDENT}

\author[a]{Hengbin~Shao,\footnote[2]{\label{co_author}These authors contributed equally to this work.}}
\author[a]{Fuyudi~Zhang,\textsuperscript{\ref{co_author}}}
\author[a]{Qichao~Chang,}
\author[a]{Shuhua~Hao,}
\author[a]{Ruike~Cao,}
\author[a]{Jingtao~Huang,}
\author[a]{Weilun~Huang,}
\author[a]{Hai~Liu,}
\author[a,b,c]{Hualin~Mei,}
\author[a]{Iwan~Morton-Blake,\textsuperscript{\ref{co_correspondance}}}
\author[a]{Wei~Tian,}
\author[a]{Yingwei~Wang,}
\author[a,b,c]{Xin~Xiang,\textsuperscript{\ref{co_correspondance}}}
\author[a,b,c]{Donglian~Xu,\footnote[1]{\label{co_correspondance} Corresponding authors}}

\affiliation[a]{State Key Laboratory of Dark Matter Physics, Tsung-Dao Lee Institute, Shanghai Jiao Tong University, Shanghai 201210, China}
\affiliation[b]{School of Physics and Astronomy, Shanghai Jiao Tong University, Key Laboratory for Particle Astrophysics and Cosmology (MoE), Shanghai Key Laboratory for Particle Physics and Cosmology, Shanghai 200240, China}
\affiliation[c]{Hainan Research Institute, Shanghai Jiao Tong University, Hainan 572024, China}

\emailAdd{iblake@sjtu.edu.cn, xxiang@sjtu.edu.cn, donglianxu@sjtu.edu.cn}

\abstract{

TRIDENT is a planned multi-cubic-kilometer deep-sea neutrino telescope to be built in the South China Sea, designed to rapidly discover high-energy astrophysical neutrino sources with sensitivity to all neutrino flavors. Achieving this at scale requires a detector design that balances performance with power, cost, and mechanical simplicity. This study presents a cost-effective optimization of TRIDENT's hybrid Digital Optical Module (hDOM) design, comparing configurations using high-quantum-efficiency (QE) 3-inch PMTs and larger 4-inch PMTs, the latter evaluated with both baseline and enhanced QE assumptions. Using full-chain detector simulations incorporating site-specific seawater optical properties and realistic backgrounds, we assess performance in all-flavor neutrino detection efficiency, directional reconstruction, and $\nu_\tau$ flavor identification from 1~TeV to 10~PeV. We find that if 4-inch PMTs can achieve QE comparable to 3-inch PMTs, their performance matches or improves upon that of the 3-inch design, while significantly reducing channel count, power consumption, and cost. These findings support the 4-inch PMT hDOM as a promising and scalable choice for TRIDENT’s future instrumentation.

}


\begin{document}

\maketitle


\flushbottom

\section{Introduction}
The origin of highly energetic cosmic rays is a question that has been left unanswered for over a century. Neutrino telescopes are pivotal instruments in astrophysics and particle physics, enabling the detection of neutrinos from potential cosmic ray sources such as active galactic nuclei, blazars, gamma-ray bursts and supernovae~\cite{IceCube:2022agn,IceCube:2018dnn,IceCube:2017zya,Scholberg:2012id}. These observations provide unique insights into cosmic ray acceleration mechanisms, high-energy astrophysical processes and fundamental particle interactions. Building next-generation neutrino telescopes with instrumented volumes on the scale of cubic kilometers necessitates cost-effective and scalable detector designs. A crucial component of these detectors is the photomultiplier tube (PMT), which converts faint Cherenkov light signals from neutrino interactions into electrical pulses, forming the backbone of the detection system.

The TRopIcal DEep-sea Neutrino Telescope (TRIDENT) is an envisioned next-generation neutrino telescope designed to rapidly discover multiple astrophysical neutrino sources, with strong sensitivity to neutrinos of all flavors. TRIDENT plans to instrument several cubic kilometers of deep ocean water in the South China Sea with advanced optical detection modules. Its near-equatorial location allows for observation of the entire sky as the Earth rotates, complementing IceCube and northern-hemisphere detectors~\cite{trident2023na}. The detector expects to have first-rate neutrino direction resolution and flavor sensitivity across an extended energy range. To achieve this, TRIDENT employs hybrid Digital Optical Modules (hDOMs), which include fast-responding Silicon photomultipliers (SiPMs) alongside PMTs, all contained within a pressure-resistant glass vessel.

While first-generation detectors (IceCube, ANTARES, Baikal-GVD) employed a single large PMT in each DOM~\cite{icecube:2006dom, ANTARES:2005, Baikal:2016}, the field has moved toward multi-PMT DOM (mDOM) design, where each module contains an array of smaller PMTs, currently used by KM3NeT \cite{KM3NeT:2022pnv} and expected to be employed in IceCube-Gen2~\cite{IceCube2:2021} and P-ONE~\cite{Agostini:2019wsd}. This advancement offers a number of advantages. The array of small PMTs are arranged to view nearly $4\pi$ solid angle, providing a homogeneous angular coverage and a greatly increased total photocathode area compared to a single PMT module. The spatial distribution of PMTs enables intrinsic directional sensitivity - each hit PMT effectively provides  pointing information. Having multiple independent PMT channels enables the application of local coincidences within a DOM, which allow for the suppression of uncorrelated single-photon noise e.g. dark count rate and radioactive backgrounds, by requiring coincident hits on several tubes. These advantages have been demonstrated in practice: the KM3NeT mDOM, for instance, achieves effective background rejection and improved photon directionality, validating the multi-PMT approach for large neutrino arrays~\cite{km3net:2016}. 

While the mDOM architecture represents the future, the choice of PMT model and configuration within each module remains a critical factor influencing overall photon detection sensitivity and resolution. As of the time of writing, leading manufacturers such as Hamamatsu Photonics and North Night Vision Technology (NNVT) offer a range of small-diameter PMTs suitable for deep-sea and deep-ice deployment at scale. High quantum efficiency (QE) is essential for maximizing photon detection efficiency. Hamamatsu’s lineup includes high QE models such as the R14688-100 which achieves QE values exceeding 30\% near 400~nm wavelength~\cite{hamamatsu_r14688,kaptanoglu2024r14688}. NNVT, known for developing the 20-inch MCP-PMT used in JUNO, has also produced 3-inch and 4-inch PMTs with comparable high QE performance~\cite{IceCube-Gen2:2023qpa}. Another key parameter is the transit time spread (TTS), which defines the timing resolution of a PMT and is crucial for reconstructing the arrival direction of Cherenkov light fronts—an essential capability for identifying and classifying neutrino events. Both companies produce PMTs with single photoelectron TTS less than 3~ns.


The TRIDENT collaboration has developed a prototype hDOM including 3-inch PMTs and SiPM units~\cite{fanicrc2021, trident2023hdom}, which will be deployed in TRIDENT Phase-1. In expanding hDOM production to TRIDENT's full array, a next-generation scale of neutrino telescope, considerations of power consumption and cost must be carefully weighed against expected detector performance. This work addresses a cost effective optimization of candidate hDOMs designs, evaluating the use of 3-inch and 4-inch PMT hDOM configurations.

\section{Design Configurations}

This study considers two preliminary hDOM designs based on a standard 17-inch outer diameter pressure vessel with a 14 mm wall thickness, manufactured by Nautilus. The 17-inch vessel is a relatively cost-effective solution with over two decades of demonstrated reliability in deep-sea neutrino telescope operations~\cite{ANTARES:2025exu, KM3NeT:2022pnv}. The vessel consists of two hemispherical glass shells and is pre-equipped with a vacuum port for evacuation and sealing, along with a penetrator feedthrough for power and data transmission, both located at the top hemisphere. The photosensors are mounted using aluminum alloy support structures that provide mechanical stability and allow accurate orientation of each sensor. A transparent optical gel is filled between the PMT windows and the inner surface of the sphere to minimize photon loss across refractive index boundaries. Since convective cooling is absent in the sealed low-pressure environment, the heat generated by the electronics is channeled to an umbrella-shaped cap at the top, thermally coupled to the glass vessel, which conducts heat to the surrounding seawater. These shared mechanical and environmental constraints define the boundary conditions for the two designs considered in this work.

Within this 17-inch enclosure, we evaluate two configurations that differ in PMT size and number. Both designs adopt the multi-PMT concept to maximize $4\pi$ optical coverage, but differ in their geometry and segmentation. Each design is developed within the same mechanical and environmental constraints outlined above.

\subsection{3-inch PMT hDOM Design}
\begin{figure}[h!]
    \centering
    \begin{subfigure}{0.43\textwidth}
        \centering
        \includegraphics[width=\linewidth]{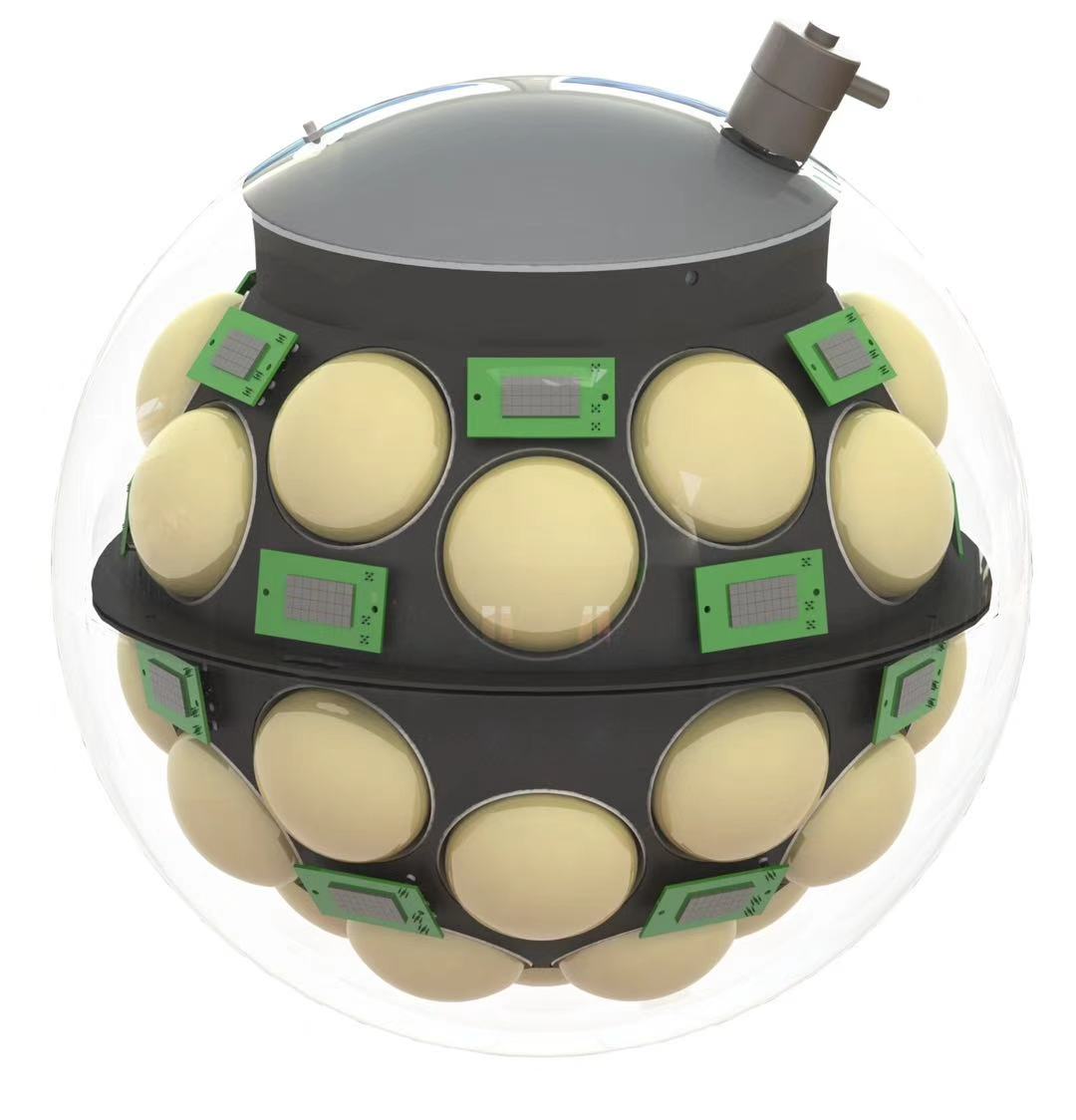}
        \subcaption{}
        \label{fig:3inchDOM}
    \end{subfigure}
    \begin{subfigure}{0.43\textwidth}
        \centering
        \includegraphics[width=\linewidth]{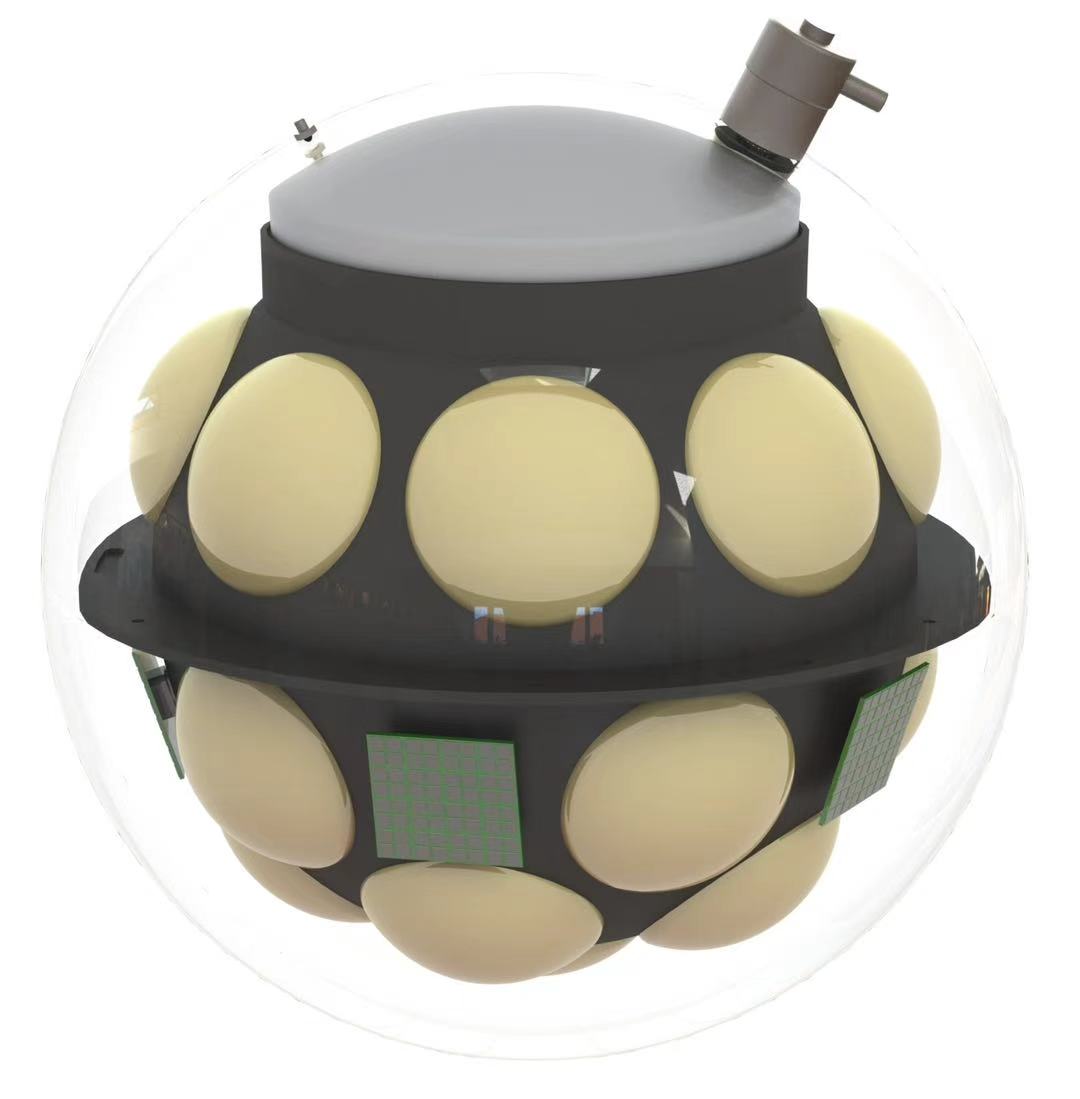}
        \subcaption{}
        \label{fig:4inchDOM}
    \end{subfigure}
    \caption{Illustrations of the 3-inch and 4-inch PMT hDOM prototype designs tested in this work. (a) 3-inch PMT hDOM containing 31 PMTs and 24 SiPM arrays. (b) 4-inch PMT hDOM containing 19 PMTs and 5 SiPM arrays.}
    \label{fig:hDOMDesigns}
\end{figure}

The first configuration employs an array of thirty-one 3-inch PMTs, following the same geometry outlined in previous TRIDENT publications~\cite{trident2023hdom}, where Fig.~\ref{fig:3inchDOM} shows a rendering of the design. The 31 tubes are arranged in five concentric rings with zenith angles of 55\textdegree, 73\textdegree, 107\textdegree, 125\textdegree, and 150\textdegree, respectively. Each ring contains six PMTs spaced evenly at 60\textdegree intervals in azimuth. Adjacent rings are azimuthally offset by 30\textdegree relative to one another, so that the PMTs in one ring face the gaps between PMTs in the next ring, avoiding blind spots. The final PMT is placed at the bottom of the module, pointing directly downward at a zenith angle of 180\textdegree.

\begin{figure}[h!]
    \centering
    \begin{subfigure}{0.4\textwidth}
        \centering
        \includegraphics[width=\linewidth]{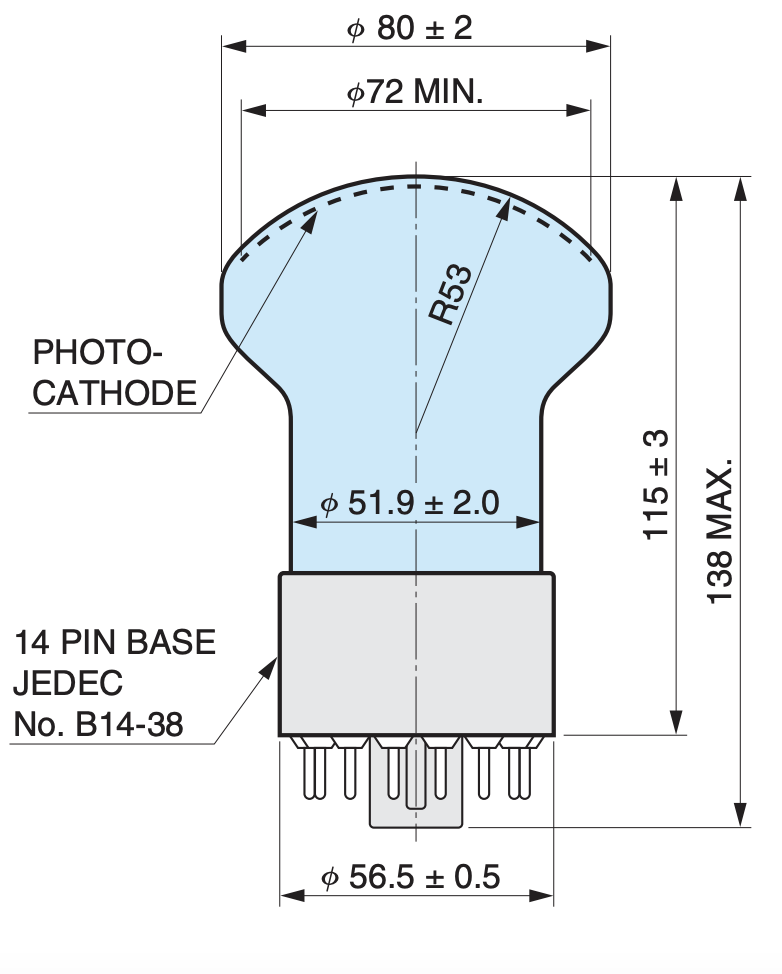}
        \subcaption{}
    \end{subfigure}
    \begin{subfigure}{0.5\textwidth}
        \centering
        \includegraphics[width=\linewidth]{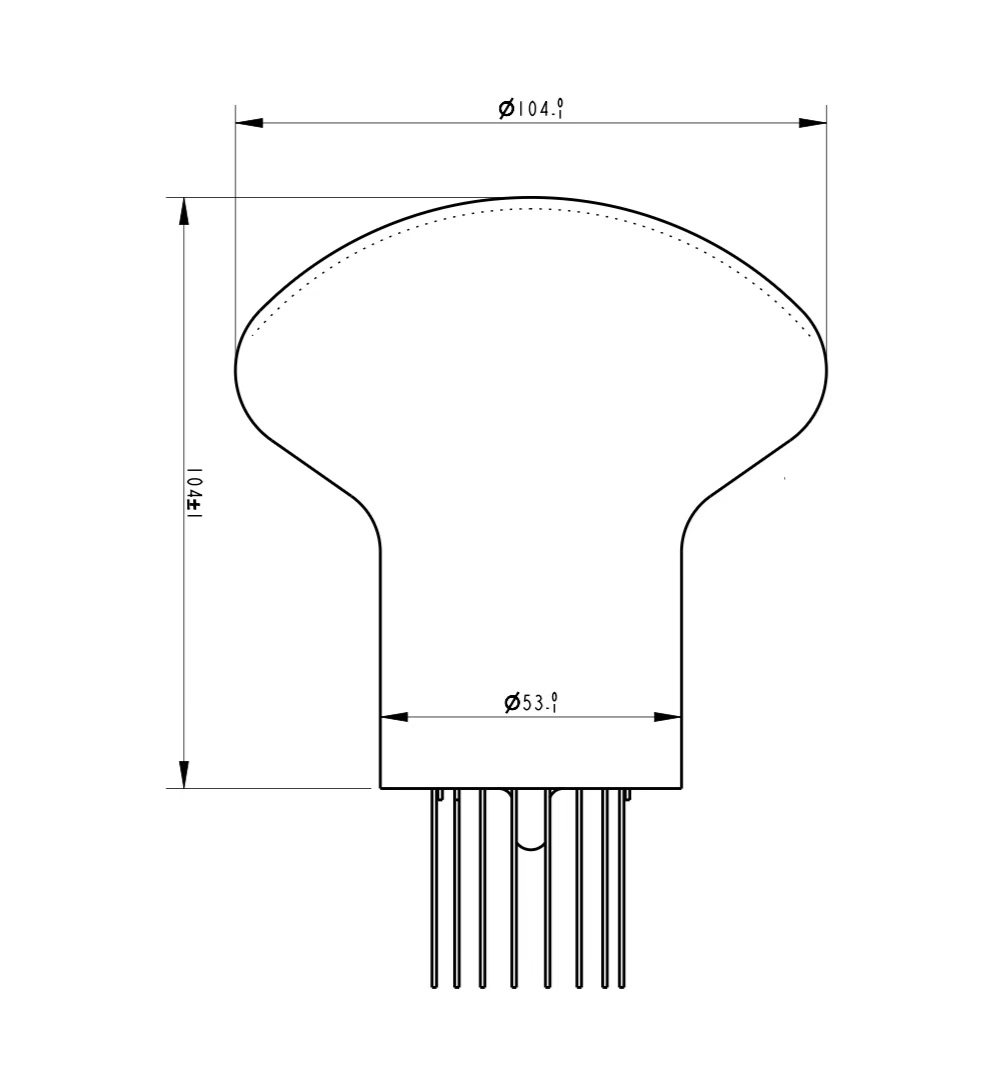}
        \subcaption{}
    \end{subfigure}
    \caption{Blueprints of PMT dimensions for (a) 3-inch PMT Hamamatsu Model R14374 and (b) 4-inch PMT NNVT Model N2042.}
    \label{fig:PMTDimensions}
\end{figure}

The PMT used is the Hamamatsu R14374 model, which is a 3-inch PMT with a circular bialkali photocathode and a 10-stage dynode structure, shown in Fig.~\ref{fig:PMTDimensions}. This compact PMT offers high quantum efficiency (QE), fast timing, and stable gain, and has been evaluated for the KM3NeT Phase-2.0 upgrade~\cite{KM3NeT:2025pmt}. In this work, we adopt a slightly shortened variant of the R14374, developed in collaboration with the manufacturer, featuring enhanced QE and reduced TTS. This variant operates at similar bias voltages (typically 1000–1400V) and has been tested in prototype 3-inch hDOM assemblies. QE curves and TTS performance were provided by Hamamatsu and used as input in simulation performance studies, where key PMT parameters are summarised in Tab.~\ref{tab:PMTtable}. A prototype motherboard has been developed to read out waveforms from the 31 PMTs and time stamps from the 24 SiPM arrays~\cite{trident2024sipm}. Current power consumption for the hDOM is an ongoing area of development, including optimization of the motherboard \cite{yangyong2025}. This may also be simply reduced by employing a smaller number of channels per hDOM. Introduced in the following section is a candidate hDOM design containing 4-inch PMTs.

\begin{figure}[h!]
    \centering
        \includegraphics[width=0.7\linewidth]{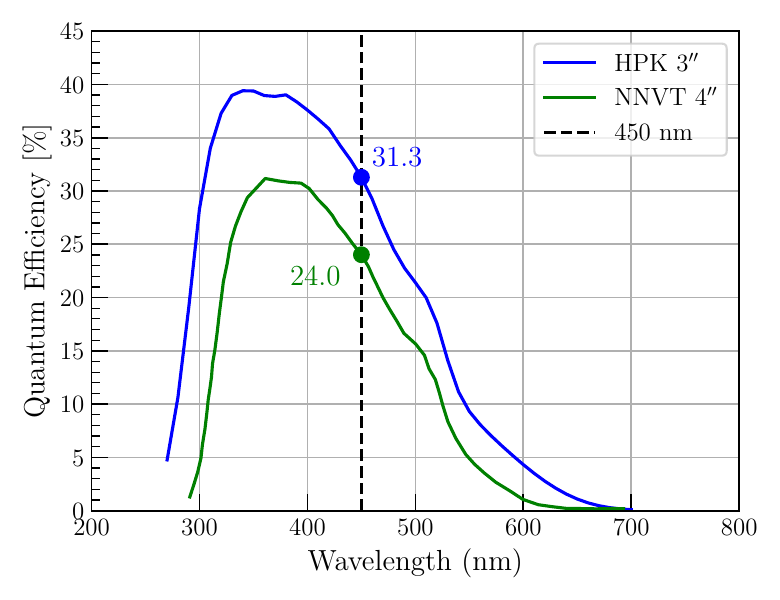}
        \caption{Quantum efficiency (QE) measurements as a function of incident photon wavelength adapted from manufacturers, showing the 3-inch and 4-inch PMT models listed in table \ref{tab:PMTtable}. The 4-inch distribution is adapted from IceCube-Gen2 measurements~\cite{IceCube-Gen2:2023qpa}. Throughout this work the blue and green QE distributions are referred to as High QE and Low QE, respectively.}
    \label{fig:PMTQE}
\end{figure}

\subsection{4-inch PMT hDOM Design}

The second configuration features nineteen 4-inch PMTs arranged within the same 17-inch pressure vessel, seen in Fig.~\ref{fig:4inchDOM}. This design reduces the number of sensors while maintaining a comparable total photocathode area to the 3-inch configuration. The PMTs are distributed in three concentric rings. The upper hemisphere contains a ring of 8 PMTs at moderate zenith angles. The lower hemisphere houses two rings, each containing 5 PMTs, positioned at steeper zenith angles. A single PMT is installed at the bottom of the module, pointing directly downward. This layout results in a greater fraction of the total photocathode area being oriented in the downward direction—a favorable feature for upward-going neutrinos that traverse the Earth and interact near the detector, expected in high-energy neutrino telescopes. 

The 4-inch hDOM geometry is based on the NNVT N2042 PMT model shown in Fig.~\ref{fig:PMTDimensions}. Two scenarios will be considered for the 4-inch. In the first baseline scenario, we adopt the QE and TTS parameters from the NNVT N2041 PMT (the predecessor of N2402), which was characterized and published by IceCube for the future Gen-2 detector. In the second hypothetical scenario, we explore an improved 4-inch PMT design with QE performance matching the high-QE Hamamatsu 3-inch PMT used in Section 2.1. This scenario serves as a benchmark for evaluating the potential physics performance of future high-performance 4-inch tubes.

\subsection{Discussion}

\subsubsection{SiPMs in hDOM designs}

hDOMs incorporate large-area SiPM arrays to enhance timing precision and photon detection. TRIDENT has developed custom front-end electronics that optimize SiPM performance, using a series–parallel connection scheme and low-noise pre-amplifier architecture to mitigate the increased capacitance from large detection areas. These SiPM arrays have demonstrated approximately 300 ps ( 600 ps) FWHM single-photon time resolution for $4\times4$ ($8\times8$) Hamamatsu S13360 SiPM arrays while maintaining modest power consumption of $\sim$90 mW ( $\sim$185 mW) per array during laboratory validation~\cite{trident2024sipm, trident2025sipm}. The fine timing resolution of SiPMs expects to aid in the reconstruction of neutrino event direction, vertex and particle discrimination. The addition of SiPMs can also fill the space between the PMTs with additional areas of photosensitivity.
Additionally, the SiPM arrays can be used for calibration, leveraging $^{40}$K decays in seawater which produce Cherenkov light that can simultaneously hit multiple sensors within the same hDOM. This enables the measurement of relative photon arrival time differences between the two sensors via coincident hits. SiPMs serve as an ideal reference due to their negligible transit time and excellent resolution, allowing precise long-term monitoring of the time jitter for PMT and front-end electronics \cite{SIMON201985}. 
A dedicated calibration strategy for TRIDENT is under development, to be introduced in a future publication.

A noteworthy difference between the 3-inch and 4-inch hDOM designs is the accommodation of SiPM arrays around the PMTs. The 4-inch design has a 2.4 times reduced total SiPM array area, where individual arrays are $8\times8$ in size instead of $4\times8$ as in the 3-inch PMT hDOM. Smaller SiPM arrays offer finer intrinsic timing resolution; however, advances in connection schemes have significantly improved the timing performance of larger arrays~\cite{trident2025sipm}. 

It has been observed in seawater experiments that DOM performance may degrade over time due to sedimentation collecting on the upper hemisphere~\cite{Antares:2002biofouling,km3_sedimentation}. To optimize their cost-effectiveness in the 4-inch PMT hDOM and combat against the overall reduced available space between PMTs, SiPMs are attached in the lower hemisphere alone, also maximizing sensitivity to upward-going neutrino events.

\subsubsection{Cost, Power, and Other Considerations}

The manufacturing process for small PMTs is broadly similar, and the cost of raw materials (e.g. glass and photocathode coatings) accounts for a modest fraction. As a result, the cost per unit photocathode area tends to remain relatively consistent across similarly sized tubes. In practice, the final cost is more strongly influenced by manufacturing yield. For reference, 3-inch PMTs developed for KM3NeT achieved a lower cost per photocathode area than traditional 10-inch tubes~\cite{Nakayama2016}. A 4-inch PMT is therefore expected to be only marginally more expensive than a 3-inch on a per-area basis. Since the 4-inch design reduces the PMT count by approximately 40\%, it also lowers the cost of associated components, such as high voltage supply, readout electronics, and connectors, which scale approximately with channel count. The 4-inch design with fewer mechanical mounts and electrical connections, results in a simplified assembly that streamlines production and enhances overall system reliability. 


Power consumption is another critical consideration for neutrino telescopes, as delivering power over long subsea distances poses significant engineering challenges. A lower per-module power draw reduces the burden on the power distribution network. The 4-inch design lowers the analog front-end and ADC power budget by reducing the number of PMTs. Although the total savings are less than the 40\% decrease in channel count due to overhead (FPGA, communication, DCDC converters), the decrease of channel number can still lead to significant power reduction. 

Beyond these considerations, the impact of these design choices on physics performance must be evaluated carefully, as explored in the following section.



\section{Impact of hDOM Design on Physics Performance}

Given the cost and power consumption differences between design configurations, it is essential to evaluate how each hDOM architecture affects key detector performance metrics relevant to neutrino detection and reconstruction across all flavors. The primary distinctions between the 3-inch and 4-inch hDOM designs stem from differences in photocathode coverage, photon detection efficiency, and timing resolution, each of which can influence physics performance.

Performance comparisons are conducted using the TRIDENTSim full detector simulation framework \cite{chang_trident_geometry_2023,zhang_trident_shower2023}, based on Geant4 and CORSIKA8 \cite{geant4_2003,hahn_corsika8_2019}. This framework incorporates detailed modeling of particle production, propagation, and light emission in seawater using measured optical properties \cite{trident2023na}, as well as the TRIDENT detector response. The main performance categories evaluated are:
\begin{itemize}
    \item hDOM trigger rates from natural background sources.
    \item Detection efficiency and angular resolution of track-like events from $\nu_\mu$ charged-current (CC) interactions.
    \item Detection efficiency and angular resolution of cascade-like events from $\nu_e$ CC interactions.
    \item Double-pulse identification efficiency for $\nu_\tau$ CC events.
\end{itemize}

\begin{table}[h!]
    \centering
    \caption{Summary of key parameters between 3 and 4-inch PMTs from Hamamatsu and NNVT~\cite{IceCube-Gen2:2023qpa}, respectively, used in simulation.} 
    \begin{tabular}{|c|c|c|}
    \hline
       Parameter & \thead{Hamamatsu \\ R14374} & \thead{NNVT \\ N2041} \\ 
       \hline
       Photocathode Diameter [cm] & 7.2 & 9.8 \\
       \hline
       Transit time spread (FWHM) [ns] & 1.4 & 2.7 \\
       \hline
       Number of PMTs per hDOM & 31 & 19 \\
       \hline
       Total photosensitive area [cm\textsuperscript{2}] & 1455 & 1704 \\
       \hline
       \end{tabular}
    \label{tab:PMTtable}
\end{table}


\subsection{Simulated hDOM Configurations}
\label{sec:hDOMDesigns}

Section~\ref{sec:hDOMDesigns} introduced two candidate hDOM configurations based on 3-inch and 4-inch PMTs. Photon detection depends on both total photocathode area and intrinsic PMT properties, such as quantum and collection efficiency. To decouple these effects and allow future performance scaling with potential PMT improvements, three distinct hDOM configurations are considered in this study:

\begin{itemize}
    \item \textbf{3-inch PMT hDOM}: Comprises of 31 Hamamatsu R14374 3-inch PMTs (Fig.~\ref{fig:3inchDOM}), with corresponding High QE shown in Fig.~\ref{fig:PMTQE} and TTS (FWHM) of 1.4~ns is used.
    \item \textbf{High QE 4-inch PMT hDOM}: Includes 19 NNVT N2042 4-inch PMTs (Fig.~\ref{fig:4inchDOM}), assuming high quantum efficiency matching the 3-inch PMT, to account for improvement forecast of 4-inch PMT models. A TTS of 1.4~ns is used.
    \item \textbf{Low QE 4-inch PMT hDOM}: Identical to the above 4-inch configuration, but assumes the  quantum efficiency and TTS of the NNVT N2041 PMT as shown in Fig.~\ref{fig:PMTQE} and Tab.~\ref{tab:PMTtable}.
\end{itemize}

These three designs are compared in terms of neutrino detection and reconstruction performance in the following sections. It should be noted that the angular acceptance of PMTs has not been taken into account in this simulation work, but expects to be included in future studies.

\subsection{hDOM Trigger Rates from Backgrounds}

To suppress triggers caused by high-rate backgrounds such as PMT dark noise and natural radioactivity in seawater, individual hDOMs employ a local coincidence-based trigger that requires multiple PMT hits within a short time window. This approach significantly reduces the impact of uncorrelated backgrounds, while maintaining high efficiency for genuine neutrino-induced signals.

The main sources of background considered are:

\begin{itemize}
    \item \textbf{Radioactivity in Seawater and Glass:} The dominant radioactive background arises from the $\beta$ decay of $^{40}$K (Q = 1.3~MeV), abundant in seawater and also present in the glass enclosures of hDOMs. Due to the low energy of the decay, resulting Cherenkov photons typically hit only a few PMTs. The assumed concentrations are 10.8~Bq/kg in seawater~\cite{trident2023na} and 61.0~Bq/kg in glass~\cite{Hannen2020}.
    
    \item \textbf{PMT Dark Noise:} Each PMT exhibits a baseline rate of thermal or electronic noise hits. A value of 1~kHz per PMT is assumed, consistent with current device specifications.
    
    \item \textbf{Atmospheric Muons:} Despite TRIDENT’s depth of $\sim$3.5~km, a residual flux of atmospheric muons remains~\cite{trident2025muslab}. These muons produce extended Cherenkov light profiles detectable across multiple hDOMs. Simulations of the muon spectrum and rate is performed with MUPAGE code~\cite{mupage2009} and account for site-specific geometry, averaging the muon flux over the full depth range of the detector. The single hDOM response is estimated accordingly.
\end{itemize}

\begin{figure}[h!]
    \centering
    \includegraphics[width=0.7\linewidth]{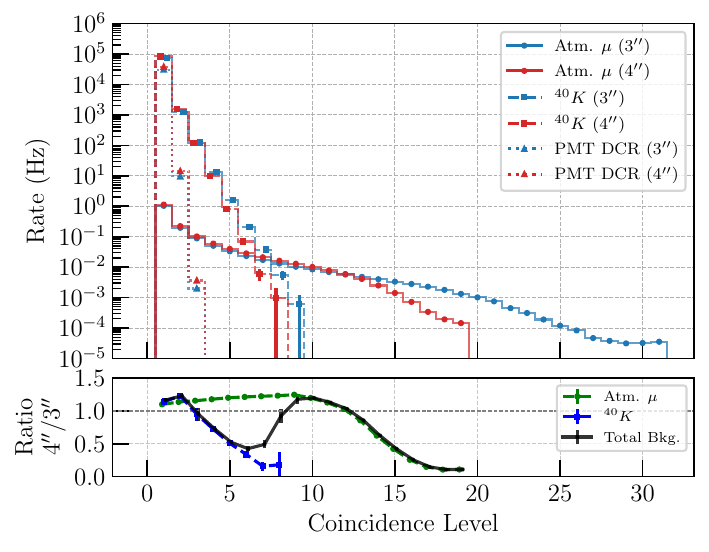}
    \caption{Expected hDOM trigger rates due to PMT dark noise, $^{40}$K radioactivity, and atmospheric muons as a function of PMT coincidence level (CL), for both the 3-inch and High QE 4-inch PMT designs. The lower panel shows the ratio of 4-inch to 3-inch PMT hDOM CL rates. Uncertainty bars reflect simulation statistics.}
    \label{fig:CLRates}
\end{figure}

The coincidence level (CL) is defined as the number of PMTs registering hits within a 20~ns time window. Fig.~\ref{fig:CLRates} compares the expected background-induced CL distributions for the 3-inch and High QE 4-inch hDOM designs, where each CL value includes hit PMT counts greater or equal to the given CL value. As expected, low-energy backgrounds (dark noise and $^{40}$K) fall off rapidly with increasing CL, while atmospheric muons produce higher CL values due to their long paths and comparatively higher light yield. $^{40}$K is shown only for seawater in Fig.~\ref{fig:CLRates} since the hDOM glass is expected to yield a sub-percent contribution to the trigger rate. The rate ratios show that the $^{40}$K-induced CL=1 rates, occurring isotopically around the hDOM, are 14$\%$ higher for the 4-inch PMTs compared to the 3-inch, reflecting the larger total photocathode surface area of the 4-inch design, summarized in Tab.~\ref{tab:PMTtable}. At higher CL, the 4-inch hDOM rates drop more steeply due to fewer PMTs per module, despite the larger photocathode size. For atmospheric muons the hit rate is 10$\%$ higher for the 4-inch design. This excess is less than the rate surplus seen for $^{40}$K. This is due to the 4-inch hDOM design having a fractionally higher photocathode coverage on its downward-facing hemisphere. Since atmospheric muons are dominantly down-going, this asymmetry affects the overall detection efficiency and alters the CL ratio behavior compared to the more isotropic $^{40}$K decays. Muon-induced rates exhibit a flatter ratio across CL, reflecting the broader light yield and the effect of hDOM angular photocathode coverage. 

\subsection{Neutrino Detection Efficiency}

Building upon the hDOM-level background suppression results, we now compare the neutrino detection efficiency of each hDOM design using the effective area, a standard performance metric in neutrino telescopes. The effective area $A_\text{eff}(E_\nu)$ represents the equivalent area over which the detector is fully efficient to an incident neutrino flux of energy $E_\nu$, and depends on the neutrino interaction cross-section, detector geometry, photon collection efficiency, and triggering performance.

The simulations presented in this section adopt the TRIDENT detector geometry defined in ~\cite{trident2023na}, comprising of a 1200-string array distributed over approximately 10~km$^3$. Each string hosts 20 hDOMs, vertically separated by 30~m, and the strings are arranged in a Penrose tiling pattern, with an average horizontal spacing of 100~m. The detector is assumed to be deployed at a depth of 3.5~km in the South China Sea, with light propagation simulated using seawater optical properties measured in-situ. This layout is used as the reference geometry for all performance comparisons presented below.

In this study, we compute the effective area separately for charged-current (CC) interactions of $\nu_\mu$ and $\nu_e$, corresponding respectively to track-like and cascade-like topologies~\cite{IceCube:2016}. Track events are generated by high-energy muons emitting Cherenkov light over long paths, whereas cascades arise from localized electromagnetic and hadronic showers.

To suppress backgrounds while retaining high efficiency, a detector-level trigger is applied requiring at least five hDOMs to satisfy a local coincidence of $\geq$2 separate PMTs within 20~ns, hereafter referred to as a '5L1' trigger. This configuration is found to be sufficient to reject random coincidences from $^{40}$K and dark noise while retaining sensitivity to signal events.

\begin{figure}[h!]
    \centering
    \begin{subfigure}{0.45\textwidth}
        \centering
        \includegraphics[width=\linewidth]{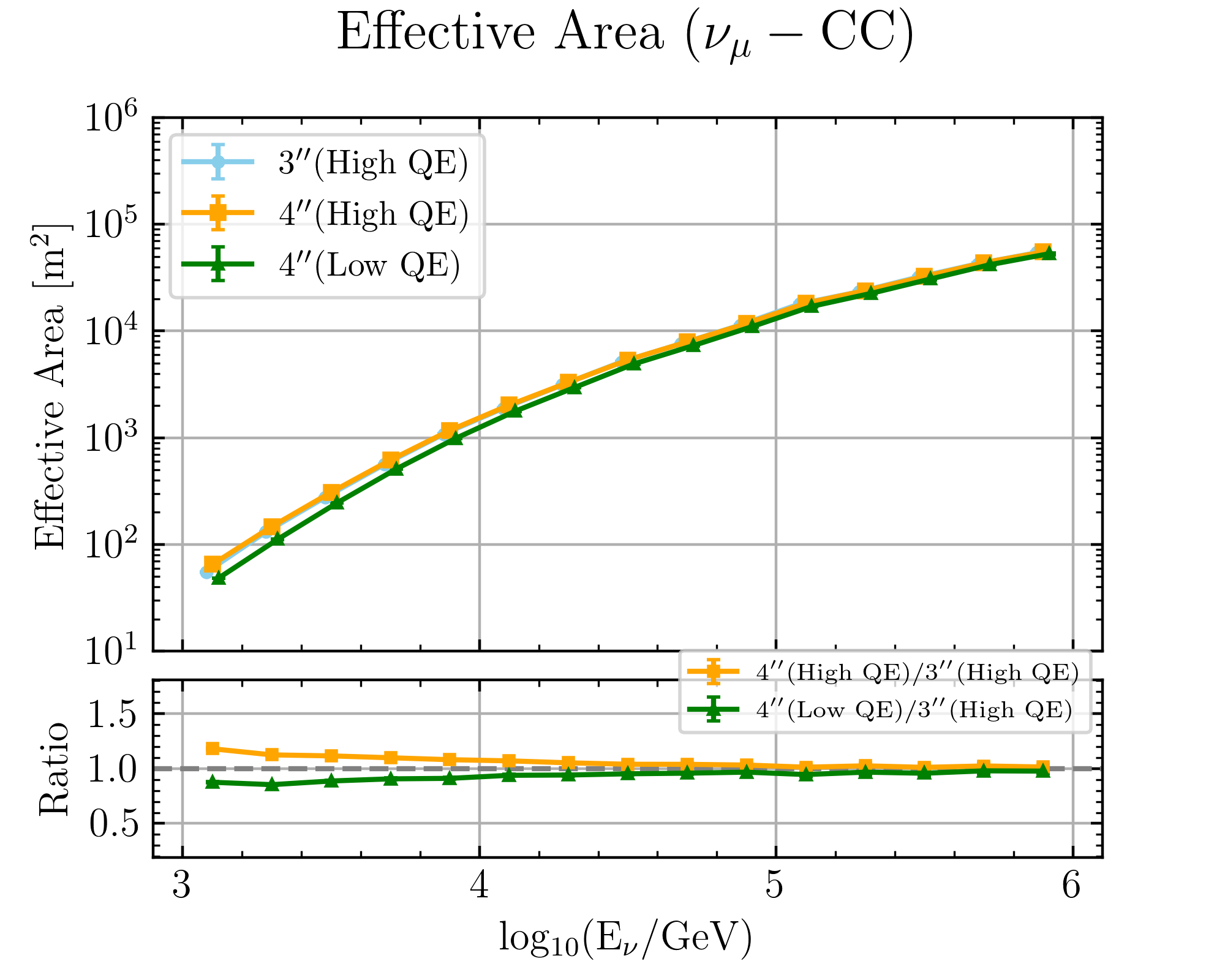}
        \subcaption{}
    \end{subfigure}
    \begin{subfigure}{0.45\textwidth}
        \centering
        \includegraphics[width=\linewidth]{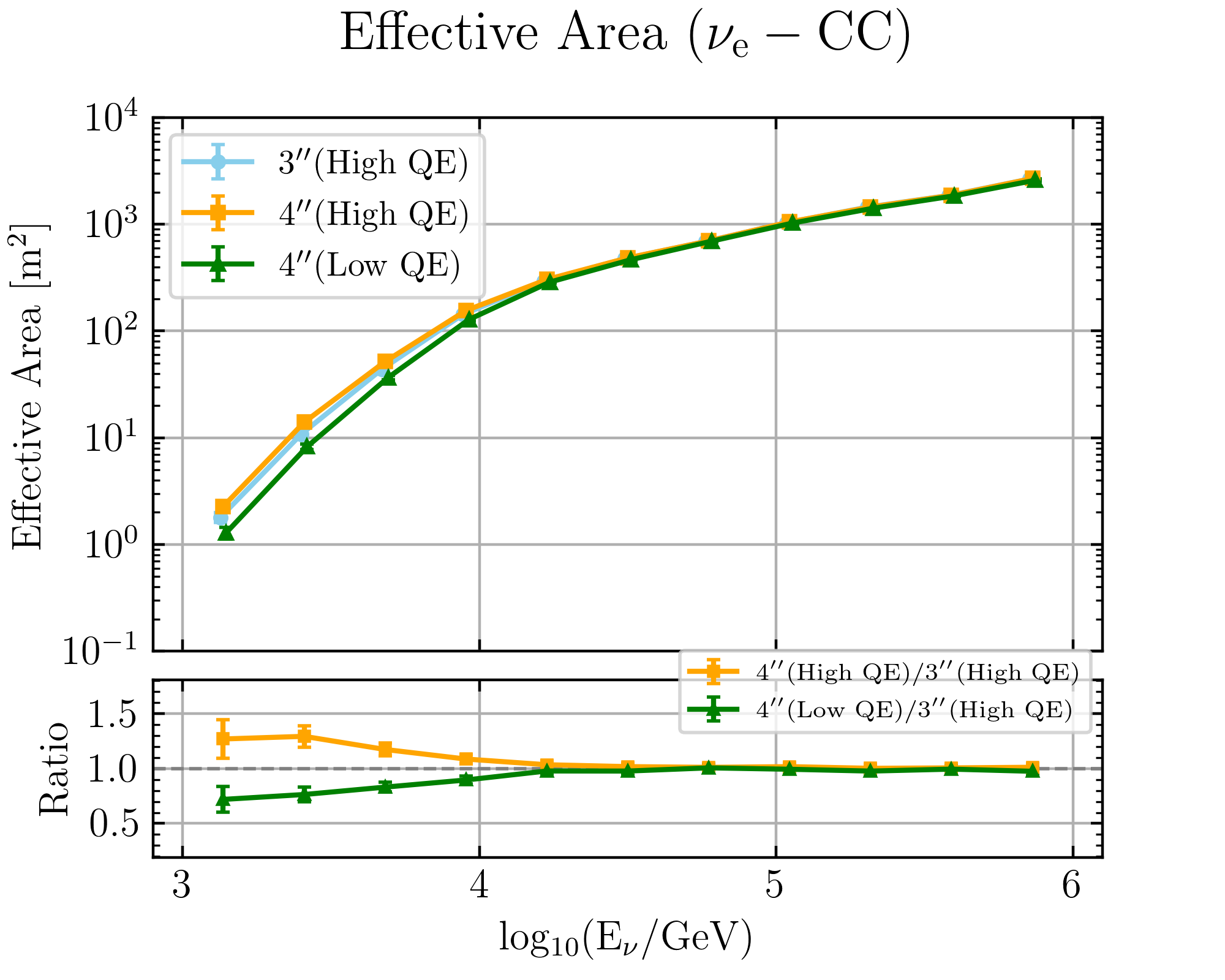}
        \subcaption{}
    \end{subfigure}
    \caption{Effective area, integrated over zenith angle, as a function of neutrino energy for (a) $\nu_\mu$ CC (track-like) and (b) $\nu_e$ CC (cascade-like) events, also showing ratios comparing 4-inch hDOM designs to the baseline 3-inch configuration. The energy range spans 1~TeV to 1~PeV.}
    \label{fig:AeffRatio}
\end{figure}

Fig.~\ref{fig:AeffRatio} compares the effective area as a function of neutrino energy for each hDOM design. Results are shown separately for $\nu_\mu$ CC and $\nu_e$ CC interactions, integrated over all zenith angles. For both interaction types, the High QE 4-inch design yields improved detection efficiency relative to the 3-inch design, particularly at lower interaction energies. These differences are primarily driven by the total photocathode coverage. The Low QE variant of the 4-inch design exhibits reduced efficiency at low energies, highlighting the importance of photon detection efficiency in threshold-limited regimes.

\subsection{Direction Reconstruction of Track and Cascade Events}

Direction reconstruction is critical for identifying astrophysical neutrino sources. The ability to resolve the incoming neutrino direction depends on the event topology and the timing granularity of recorded photon hits.

For track-like events from $\nu_\mu$ CC interactions, the muon track length provides strong directional information. Track reconstruction follows a likelihood-based method developed in ~\cite{trident2023na}, utilizing distance-dependent PMT hit time distributions as well as PMT orientation for an additional handle on the incident photon direction.

Cascade reconstruction uses a maximum likelihood method similar to that in~\cite{zhang_trident_shower2023}, based on expected photon counts per hDOM as a function of distance and angle to the vertex. PMT orientation also is incorporated to constrain the cascade vertex, enabling a comparison between hDOM designs.

As with the effective area, events are required to pass the 5L1 trigger. The angular resolution is defined as the median angle between the true and reconstructed neutrino directions, binned in true neutrino energy. In both track and cascade reconstruction methods, photons detected on SiPMs are not included in these PMT performance comparisons.

\begin{figure}[h!]
    \centering
    \begin{subfigure}{0.45\textwidth}
        \centering
        \includegraphics[width=\linewidth]{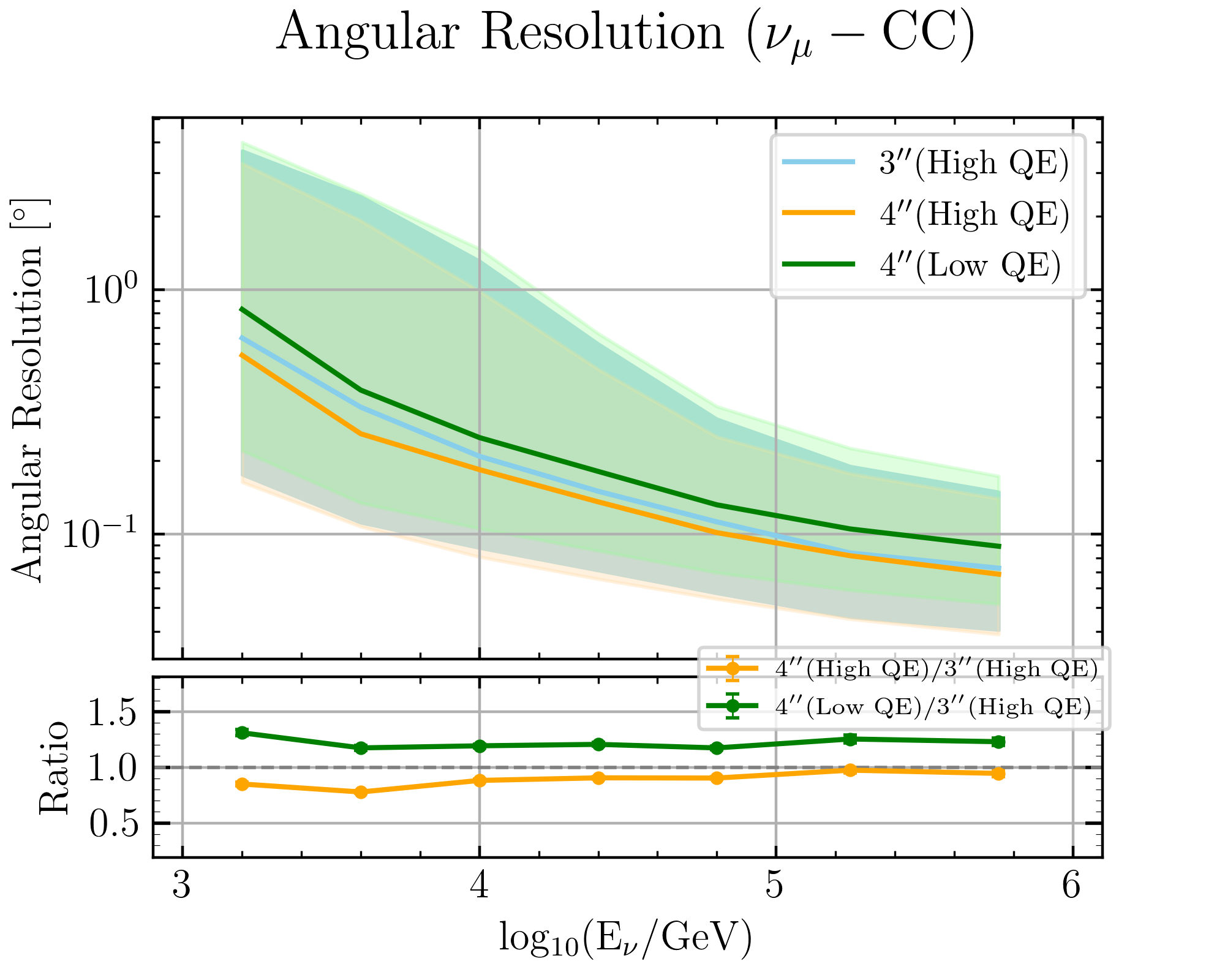}
        \subcaption{}
    \end{subfigure}
    \begin{subfigure}{0.45\textwidth}
        \centering
        \includegraphics[width=\linewidth]{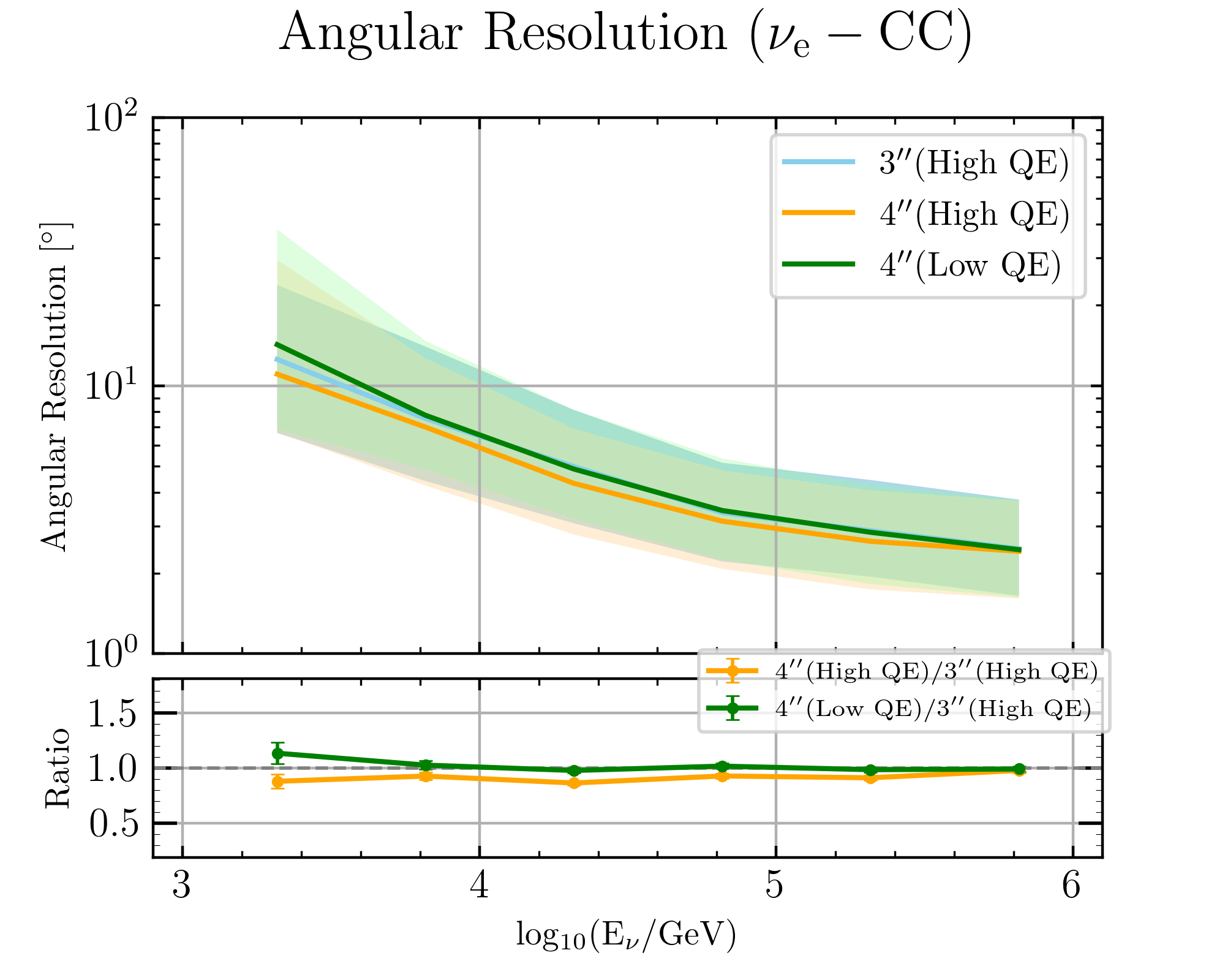}
        \subcaption{}
    \end{subfigure}
    \caption{Median angular resolution as a function of incident neutrino energy for (a) track-like $\nu_\mu$ CC events and (b) cascade-like $\nu_e$ CC events in the full TRIDENT array, also showing ratios comparing 4-inch hDOM designs to the baseline 3-inch configuration.}
    \label{fig:TrackCascadeRecon}
\end{figure}

As shown in Fig.~\ref{fig:TrackCascadeRecon}, track angular resolution improves with increasing energy, reaching $\sim0.1^\circ$ accuracy at 100 TeV energies for all designs. As with detection efficiency the 4-inch High QE design offers better resolution at lower energies due to the higher effective photon collection area. At high energies, where timing precision dominates, the resolution converges. Low QE has worsened angular resolution at all energies, indicating the importance in maximizing single PMT photon detection efficiency for accurate track reconstruction.

Cascade angular resolution evolves similarly with energy. The 4-inch Low QE design performs marginally worse at low energies, but approaches the 3-inch and High QE 4-inch configurations at higher energies due to increased light yield and sufficiently high hit statistics on hDOMs.

\subsection{Tau Neutrino Identification}

In addition to precise reconstruction of track-like and cascade-like events, TRIDENT is designed to offer enhanced sensitivity to all neutrino flavors. One important channel for flavor identification is the detection of charged-current $\nu_\tau$ interactions. At energies above TeV, $\nu_\tau$ CC events can produce a detectable, characteristic double energy deposition: one from the initial tau lepton production, and another from its subsequent decay after traveling on the order of meters. This signature can be distinguished from $\nu_e$ CC and neutral-current (NC) events, which typically appear as single, localized cascades.

TRIDENT exploits this feature using the Double-Pulse technique~\cite{IceCube:2015vkp,Tian:2023pbp,Tian:2025pbp}, which identifies two temporally separated pulses within a single hDOM waveform. When the tau decay occurs sufficiently far from the initial interaction point, and within the light-sensitive region of nearby hDOMs, the resulting delayed Cherenkov light can be detected as a second pulse in the waveform. This analysis utilizes the full-chain detector response in TRIDENTSim, including the digitization of full waveform data in each PMT, allowing time-domain identification of the two substructures. The study in this work aims to serve as an estimate of the influence hDOM design may have on the TRIDENT detector achieving one of its primary physics goals. A detailed description of the waveform selection and double-pulse identification algorithm can be found in~\cite{trident2025thesis}. A dedicated study on TRIDENT's flavor separation ability will be the focus of a future publication.

\begin{figure}[h!]
    \centering
    \includegraphics[width=0.7\linewidth]{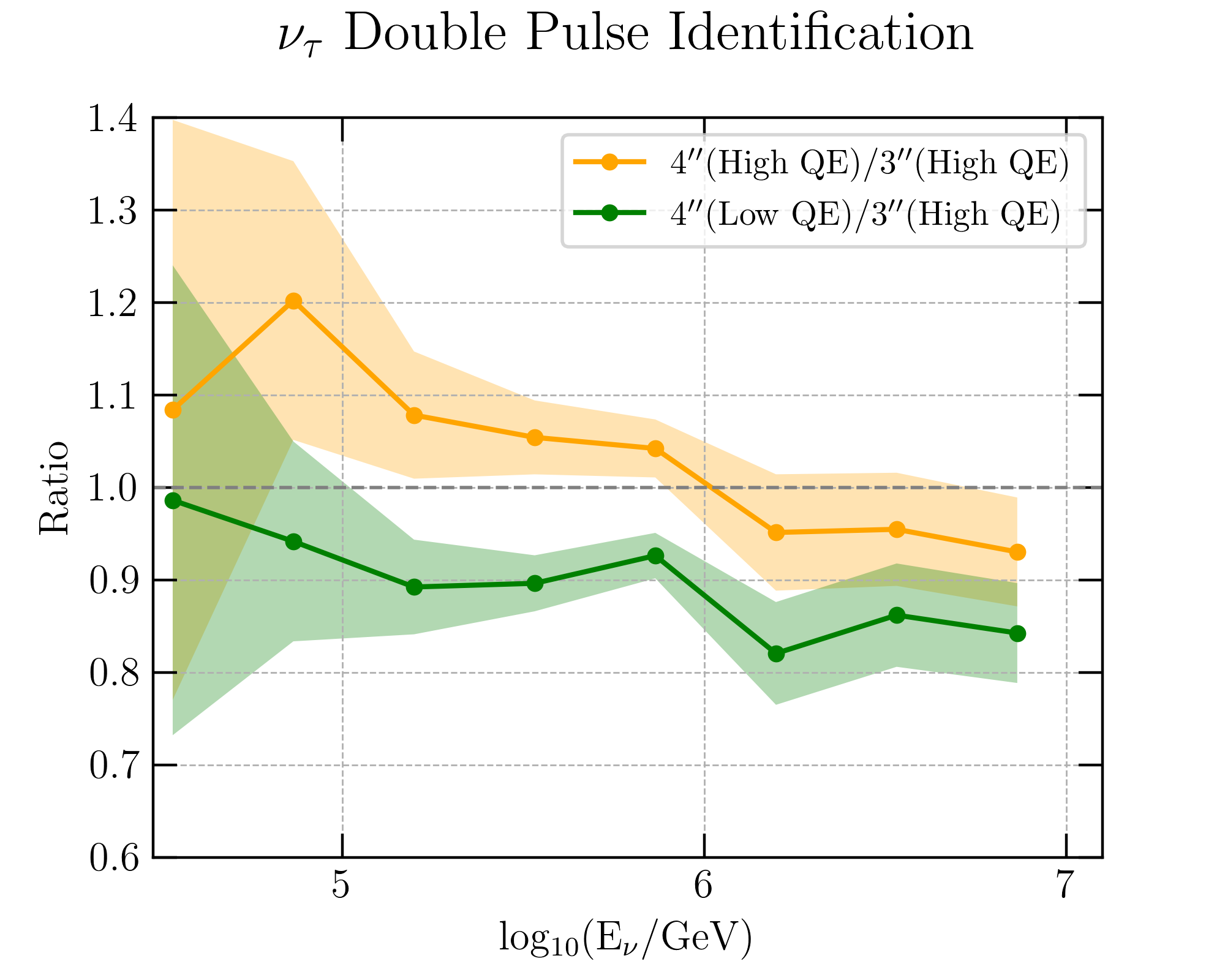}
    \caption{Ratio of $\nu_\tau$ CC double-pulse identification efficiency for the 4-inch hDOM design relative to the 3-inch design, shown as a function of neutrino energy.}
    \label{fig:DPCompare}
\end{figure}

Fig.~\ref{fig:DPCompare} shows the ratio of double-pulse identification rates for $\nu_\tau$ CC events between the 4-inch and 3-inch hDOM configurations, as a function of incident neutrino energy. This reflects the relative efficiency of each design in capturing double-pulse signatures.

It can be seen that the overall changes in detection efficiency are marginal, approximately within 10$\%$ of the 3-inch PMT design (within simulation statistical uncertainty). The $\nu_{\tau}$ detection efficiency improves or worsens based on the overall detection efficiency of each hDOM design, approximately independent of energy. At the very highest energies, a drop in efficiency of the 4-inch PMTs is seen, which potentially is related to the higher photon exposure per 4-inch PMT compared to the 3-inch, however, statistical uncertainty is significant in this region due to computation limits in simulating the highest energy $\nu_{\tau}$ events.

\subsection{Physics Performance Comparison}
This study compares two optical sensor designs for the TRIDENT neutrino telescope: one using many small 3-inch photomultiplier tubes (PMTs), and the other using fewer, larger 4-inch PMTs. We find that if the quantum efficiency of the 4-inch PMTs can be improved to match that of the 3-inch models, both designs offer very similar performance across key metrics, including background rejection, neutrino detection efficiency, and direction reconstruction. The larger PMTs enable a simpler design with fewer channels, lower power consumption, and reduced cost—making the 4-inch configuration an attractive option for TRIDENT’s future detector modules. However, performance drops significantly if quantum efficiency remains low, underlining the importance of continued development of high-efficiency 4-inch PMTs. With that improvement, the 4-inch design becomes a strong and cost-effective candidate for TRIDENT’s large-scale deployment.

\section{Conclusion and Outlook}


In this study, we conducted simulations using the TRIDENTSim framework to compare two hybrid Digital Optical Module (hDOM) configurations: a 31-PMT design using 3-inch high-QE Hamamatsu R14374 tubes and a 19-PMT design using 4-inch NNVT N2042 tubes, with both baseline and hypothetical high-QE scenarios for the latter. The simulations incorporated seawater optical properties measured in the Hai-Ling Basin, background sources ($^{40}$K decays, PMT dark noise, and atmospheric muons), and an hDOM-level level coincidence trigger, evaluating performance across neutrino flavors and energies from 1~TeV to 10~PeV.

The results reveal that the high-QE 4-inch design performs comparably to or better than the 3-inch configuration in key metrics. For background suppression, the 4-inch hDOM exhibits higher CL=1 rates from isotropic $^{40}$K noise due to its larger total photocathode area but steeper fall-off at higher coincidence levels, leading to similar overall trigger purity. Neutrino detection efficiency, quantified by effective area, shows improvements for the high-QE 4-inch design in track-like ($\nu_\mu$ CC) and cascade-like ($\nu_e$ CC) events, particularly at low energies, owing to enhanced photon collection. Angular resolution reaches $\sim$0.1\textdegree above 100 TeV for tracks in both designs, with the high-QE 4-inch variant offering better performance at TeV energies; Cascade resolution exhibits the same trend, converging at high energies. For $\nu_\tau$ CC flavor discrimination via double-pulse identification, efficiencies differ by $\sim$10\%. In contrast, the low-QE 4-inch scenario under performs, with reduced effective areas and degraded resolution at low-to-medium energies, underscoring the critical role of quantum efficiency.

The 4-inch design, with 40\% fewer channels, offers significant advantages in cost-effectiveness with lower per-unit expenses for readout components and reduced power consumption, alongside simpler mechanical integration. The impact on neutrino detection capabilities is modest if high QE is realized in 4-inch PMTs, preserving TRIDENT's goals of sub-degree angular resolution, all-flavor sensitivity, and source identification while enabling a more scalable, cost-efficient array. This positions the 4-inch hDOM as a strong candidate for TRIDENT, provided PMT advancements bridge the QE gap; otherwise, the 3-inch design remains preferable to safeguard low-energy thresholds and reconstruction quality.

We have demonstrated the benefits in cost and performance for a 4-inch PMT hDOM design, assuming their detection efficiency can match that of the more mature 3-inch models. We are actively collaborating with manufacturers such as Hamamatsu and NNVT, investigating further into the benefits and drawbacks of modular PMT designs of various sizes. This study has also adhered to the established Nautilus 17-inch VITROVEX glass vessel. Larger glass enclosures expect to allow for improved photocoverage per hDOM, however, this must be carefully balanced against reliable pressure-resistance, increased sea current drag and cost. 

\acknowledgments


This work has been supported by the Ministry of Science and Technology of China under Grant No.~2022YFA1605500 and the Office of Science and Technology of the Shanghai Municipal Government under Grant No. 22JC1410100. X. Xiang and H. Mei thank the  the Ministry of Science and Technology of China under Grant No.~2023YFC3107402 and 2023YFC3107401, respectively, as well as the Double First-Class startup funds provided by Shanghai Jiao Tong University. I. Morton-Blake also acknowledges the National Natural Science Foundation of China Grant No. 12350410499.


\bibliographystyle{JHEP}
\bibliography{biblio} 

\end{document}